\documentclass[%
reprint,
 amsmath,amssymb,
 aps,longbibliography,
]{revtex4-1}

\usepackage{natbib}
\usepackage{graphicx}
\usepackage{dcolumn}
\usepackage{bm}
\usepackage{xcolor}
\usepackage{verbatim}
\usepackage[normalem]{ulem}

\begin{document}


\title{Energy diffusion and absorption in chaotic systems with rapid periodic driving}

\author{Wade Hodson}
\email{whodson@umd.edu}

\affiliation{Department of Physics, University of Maryland, College Park, MD 20742}

\author{Christopher Jarzynski}
\email{cjarzyns@umd.edu}

\affiliation{Institute for Physical Science and Technology,
Department of Chemistry and Biochemistry, and Department of Physics, University of Maryland, College Park, MD 20742}

\date{\today}

\begin{abstract}
When a chaotic, ergodic Hamiltonian system with $N$ degrees of freedom is subject to sufficiently rapid periodic driving, its energy evolves diffusively. We derive a Fokker-Planck equation that governs the evolution of the system's probability distribution in energy space, and we provide explicit expressions for the energy drift and diffusion rates. Our analysis suggests that the system generically relaxes to a long-lived ``prethermal'' state characterized by minimal energy absorption, eventually followed by more rapid heating. When $N\gg 1$, the system ultimately absorbs energy indefinitely from the drive, or at least until an infinite temperature state is reached.
\end{abstract}

\pacs{Valid PACS appear here}

\maketitle


\section{Introduction}
\label{intro}

Time-periodic driving facilitates a rich range of classical and quantum dynamical behaviors, including synchronization and resonance \cite{vanderpolvandermark1927,chacon1994,antonsenetal2008,khassehetal2019}, localization \cite{dittrichetal1993,nagetal2014,baireyetal2017,haldardas2017}, and chaos \cite{vanderpolvandermark1927,ott2002}. Recent theoretical and experimental work has aimed to identify nonequilibrium ``phases of matter'' that might emerge in periodically driven systems \cite{khemani2016,yaoetal2017,elseetal2017}. Phenomena such as time crystallization \cite{yaoetal2017,elseetal2017,zhangetal2017,choietal2017,russomannoetal2017,yaoetal2020,kyprianidis2021} and prethermalization \cite{elseetal2017,abaninetal2017,herrmannetal2017,mori2018,morietal2018,mallayyaetal2019,howelletal2019,machadoetal2019,rajaketal2019,machadoetal2020,rubio-abadal2020,pengetal2021,kyprianidis2021} reveal that periodic driving can stabilize systems in a variety of interesting and useful states.

Energy absorption poses a potential obstacle to such stabilization of nonequilibrium states of matter. A driven open system in a nonequilibrium steady state attains a balance in which energy absorbed from the drive is dissipated into an environment, such as a thermal bath. But if a system is isolated, save its interaction with the drive, then maintaining a stable state requires the suppression of energy absorption from the drive. Much work has been devoted to understanding energy absorption, and the conditions under which it might be suppressed, in periodically driven, isolated classical and quantum systems \cite{dalessiorigol2014,lazaridesetal2014,abaninetal2015,demersjarzynski2015,ponteetal2015a,rehnetal2016,morietal2016,kuwaharaetal2016,abaninetal2017,mori2018,rajaketal2018,notarnicolaetal2018,howelletal2019,machadoetal2019,tranetal2019,rajaketal2019} . 

In this paper we study the general problem of energy absorption in isolated classical chaotic systems subject to rapid periodic driving. We argue that the energetic dynamics of such systems are diffusive, and we derive a Fokker-Planck equation for the evolution of the system's energy probability distribution, $\eta(E,t)$. The drift and diffusion coefficients in this equation, characterizing energy absorption and the spreading of the energy distribution, are given explicitly in terms of the dynamics of the undriven system -- much as in the case of ordinary linear response theory~\cite{dorfman1999}, but without the assumption of weak driving. For many-body systems our results suggest a scenario marked by three stages: initial relaxation to an equilibrium-like ``prethermal'' state \cite{elseetal2017,abaninetal2017,herrmannetal2017,mori2018,morietal2018,mallayyaetal2019,howelletal2019,machadoetal2019,rajaketal2019,machadoetal2020,rubio-abadal2020,pengetal2021}, followed by a long interval of minimal energy absorption, and finally rapid absorption toward an infinite-temperature state. 

Our description provides a comprehensive, quantitative account of energy absorption in rapidly and periodically driven chaotic systems.  It reveals how chaos in phase space facilitates stochastic energy evolution, how energy diffusion leads to the breakdown of the prethermal regime, and how energy absorption rates are determined by the underlying, undriven Hamiltonian dynamics. Our framework also suggests a generic explanation for the exponential-in-frequency suppression of energy absorption observed in a range of systems \cite{abaninetal2015,morietal2016,kuwaharaetal2016,elseetal2017,abaninetal2017,mori2018,howelletal2019,machadoetal2019,tranetal2019,rubio-abadal2020,pengetal2021}. Finally, we argue that the classical results that we obtain are relevant to energy absorption in quantum systems, in an appropriate semiclassical limit.

In Sec.~\ref{setup} we define the problem we will study. In Sec.~\ref{diffusion} we argue that the energy of a rapidly driven chaotic system evolves diffusively, and we derive the Fokker-Planck equation that describes this evolution. In Sec.~\ref{absorptionprethermal} we analyze energy absorption and prethermalization in the context of our energy diffusion model. In Sec.~\ref{quantum} we briefly consider the quantum counterpart of our classical problem, and we conclude in Sec.~\ref{conclusion}.

\section{Setup}
\label{setup}

Our object of study is a classical Hamiltonian system with $N\ge 2$ degrees of freedom. At time $t$, the microscopic state of the system is specified by a phase space point $z_t \equiv z \equiv (\mathbf{q},\mathbf{p})$, where the $N$-vectors $\mathbf{q}$ and $\mathbf{p}$ specify canonical coordinates and momenta. The system evolves under Hamilton's equations of motion, generated by a Hamiltonian $H \equiv H(z,t)$:
\begin{equation}
\label{eom}
\frac{d \mathbf{q}}{d t} = \frac{\partial H}{\partial \mathbf{p}}, \quad \frac{d \mathbf{p}}{d t} = -\frac{\partial H}{\partial \mathbf{q}}.
\end{equation}
For an ensemble of trajectories, the phase space probability distribution $\rho (z,t)$ obeys the Liouville equation
\begin{equation}
\label{liouville}
\frac{\partial \rho}{\partial t} = \{H, \rho \}
= \frac{\partial H}{\partial \mathbf{q}} \cdot \frac{\partial \rho}{\partial \mathbf{p}} -
\frac{\partial H}{\partial \mathbf{p}} \cdot \frac{\partial \rho}{\partial \mathbf{q}}
,
\end{equation}
where $\{\cdot,\cdot\}$ denotes the Poisson bracket~\cite{Goldstein80}. We take $H(z,t)$ to be a periodic function of time, with  period $T$, and we decompose this Hamiltonian into its time average $H_0 (z) \equiv T^{-1} \int_0^T dt \, H(z,t) $, and a remainder $V(z,t) = V(z,t+T)$ with vanishing average:
\begin{equation}
\label{hamiltonian}
 H(z,t) = H_0(z) + V(z,t).
\end{equation}
We will refer to $H_0(z)$ as the ``bare'' or ``undriven'' Hamiltonian, and to $V(z,t)$ as the ``drive.'' The former determines the evolution of the system in complete isolation, that is when $V=0$. We define the system's energy $E(t)$ to be the bare Hamiltonian evaluated at $z_t$: $E(t) \equiv H_0(z_t)$. When $V=0$ the energy is a constant of the motion: any trajectory $z_t$ is constrained to evolve on an {\it energy shell}, that is a level surface of $H_0(z)$. The evolution of the energy when $V\ne 0$ will be our central focus.
The magnitude of the drive $V(z,t)$ is arbitrary; in particular we do not assume it to be small.

We assume that the undriven dynamics are chaotic and ergodic on each energy shell of $H_0$. 
Such dynamics exhibit {\it mixing} as trajectories diverge from one another exponentially with time~\cite{ott2002}.
This leads to a loss of statistical dependence between states of the system at different times, as reflected in the decay of correlations -- in effect the system loses its memory of previously visited states.
After a characteristic mixing time, any smooth initial distribution on the energy shell evolves to a distribution that for practical purposes is microcanonical, or thermal~\cite{reichl1980,dorfman1999}.
Thus chaos offers a way to understand the self-thermalizing properties of many-body systems such as gases and liquids, while also providing low-dimensional analogues, such as chaotic billiard systems, that are accessible to numerical or analytical study.

We are interested in the limit of high driving frequency $\omega = 2 \pi/T$. When $\omega \to \infty$ the effect of the drive averages to zero over each period, as the system cannot appreciably react to the drive in such a short time. In this limit the evolution generated by the driven Hamiltonian approaches the undriven evolution: for given initial conditions $z_0$, and over a fixed time interval $0\le t\le\tau$, the trajectory $z_t$ that evolves under $H(z,t)$ converges, as $\omega\rightarrow\infty$, to the trajectory $z_t^0$ that evolves under $H_0(z)$~\cite{murdock1999,rahavetal2003}. This limit can be attained regardless of the strength of the drive, $V(z,t)$. In our case we assume $\omega$ is sufficiently high that driven and undriven trajectories remain close over timescales characteristic of the decay of correlations. 

These considerations lead to the following picture.
For sufficiently short times, energy is approximately conserved, and the driven trajectories $z_t$ generated by $H(z,t)$ are similar to the undriven trajectories $z_t^0$ generated by $H_0(z)$, allowing us to use the latter to estimate correlation functions that will arise in our analysis.
These correlation functions, as we shall see, will in turn describe how the system absorbs energy from the external drive on much longer timescales.

\section{Energy diffusion}
\label{diffusion}

Given the assumptions mentioned in Section \ref{setup}, we now argue that the energy of the driven system evolves diffusively. For simplicity, we consider monochromatic driving,
\begin{equation}
V(z,t) = V(z)\cos (\omega t) \, . 
\end{equation}
Our analysis can be generalized to  arbitrary time-periodic driving, by decomposing $V(z,t)$ in a Fourier series with fundamental frequency $\omega$.

\subsection{Argument for energy diffusion}
\label{argument}

To begin, we consider a system that evolves over a time interval $0\le t\le\Delta t$, from initial conditions $z_0$ sampled from a microcanonical distribution at energy $E_0$:
\begin{equation}
\label{micro}
 \rho(z_0,0) = \rho_{E_0} (z_0) \equiv \frac{1}{\Sigma (E_0)} \delta(E_0 - H_0 (z_0)).
\end{equation}
Here
\begin{equation}
\label{eq:dos}
\Sigma (E) = \frac{\partial\Omega}{\partial E} = \int dz \, \delta(H_0 (z) - E)
\end{equation}
is the classical density of states;
\begin{equation}
\Omega(E) = \int dz \, \theta(E-H_0 (z))
\end{equation}
is the phase space volume enclosed by the energy shell $E$, which we assume to be finite for all $E$; and the integrals are over phase space.
Let $\Delta E (z_0)$ denote the net change in the system's energy from $t=0$ to $t=\Delta t$. By Hamilton's equations, $\Delta E (z_0)$ is the time integral of the power
\begin{equation}
\label{power}
\frac{d E}{d t} \equiv \frac{d}{dt} H_0(z_t) = -  \cos (\omega t) \dot{V}(z_t),
\end{equation}
 where $\dot{V}(z) \equiv \{ V,H_0\}$.  The quantity $\Delta E$ can be viewed as a random variable, whose value is determined by the sampled initial conditions $z_0$. Understanding the statistics of $\Delta E$ in the high-frequency driving regime, for an appropriate choice of $\Delta t$ (to be clarified below), will be the key to establishing diffusion in energy space.

[We note in passing that if the undriven Hamiltonian has the form $H_0 = \sum_n({\bf p}_n^2/2m_n) + U_0(\{{\bf q}_n\})$
and if the drive $V$ depends on coordinates $\{{\bf q}_n\}$ but not momenta $\{{\bf p}_n\}$, then \eqref{power} becomes $dE/dt = \sum_n{\bf F}_n\cdot{\bf v}_n$, where ${\bf F}_n$ is the driving force acting on the $n$'th particle, and ${\bf v}_n$ is that particle's velocity.]

We now explicitly assume the driving is rapid.  To begin, we impose the condition
\begin{equation}
\label{eq:smallT}
T \ll \tau_C(E_0),
\end{equation}
where $T=2\pi/\omega$ is the drive period and $\tau_C(E_0)$ is a characteristic timescale over which chaotic mixing on the energy shell $E_0$ produces the decay of correlations.  Heuristically, \eqref{eq:smallT} implies that a trajectory $z_t$ travels only a negligible distance during one period of driving.  This condition produces the averaging over oscillations that (as mentioned in Sec.~\ref{setup}) results in driven trajectories $z_t$ resembling their undriven counterparts $z_t^0$.  Let us now choose $\Delta t$ so that over the interval $[0,\Delta t]$ the driven trajectories in our ensemble remain close to the initial energy shell $E_0$.  Chaotic mixing ensures then that a microcanonical distribution is approximately maintained.  Thus for any $t \in [0,\Delta t]$, the ensemble of points $z_t$ are approximately distributed according to the initial microcanonical ensemble.

With this picture in mind, let us divide the time interval $[0,\Delta t]$ into $M \gg 1$ subintervals of equal duration $\delta t = \Delta t/M$, and consider $\Delta E=\sum_i \delta E_i$ as a sum of subinterval energy changes $\delta E_i$, $i=1,2 \cdots M$. Each increment $\delta E_i$ is itself a random variable, determined by integrating the power \eqref{power} along the trajectory $z_t$ over the subinterval.  By the arguments of the previous paragraph, the $\delta E_i$'s  have nearly identical, microcanonical statistics, provided we choose $\delta t$ (and therefore also $\Delta t$) to be an integer multiple of the driving period $T$ to ensure that each subinterval begins at the same phase of the drive.

Chaotic mixing on the energy shell $E_0$ produces the decay of correlations. Let us further choose $\delta t$ to be longer than the characteristic correlation time $\tau_C(E_0)$, so that each $\delta E_i$ is approximately statistically independent from the others. The energy change $\Delta E$ is then a sum of $M \gg 1$ approximately independent and identically distributed increments $\delta E_i$: the system effectively performs a random walk on the energy axis.  By the central limit theorem $\Delta E$ is a normally-distributed random variable, whose mean and variance grow (for fixed $\delta t$) in proportion to the number of increments $M$, equivalently the time elapsed $\Delta t$. 

The statistical behavior just described is characteristic of a diffusive process in energy space, motivating us to model it by a Fokker-Planck equation \cite{gardiner1985}.  Letting
\begin{equation}
\eta(E,t) \equiv \int dz \, \delta(H_0 (z) - E) \rho(z,t)
\end{equation}
denote the energy distribution, we  postulate that the time evolution of $\eta$ is given by
\begin{equation}
\label{fp}
 \frac{\partial \eta}{\partial t} = -\frac{\partial}{\partial E} \left( g_1 \eta \right)+ \frac{1}{2} \frac{\partial ^2}{\partial E^2} \left( g_2 \eta \right) .
\end{equation}
The drift and diffusion coefficients $g_1 (E,\omega)$ and $g_2 (E,\omega)$ characterize, respectively, the rate at which the distribution $\eta$ shifts and spreads on the energy axis; see \eqref{eq:mean} and \eqref{eq:var} below.  These coefficients depend on the system energy $E$ and the driving frequency $\omega$.   Energy diffusion and its description in terms of the Fokker-Planck equation have been studied in various contexts involving externally driven Hamiltonian systems \cite{ott1979,brownetal-prl1987,brownetal1987,wilkinson1990,linkwitzgrabert1991,jarzynski1992,jarzynski1993,cohen2000,buninetal2011,debievreparris2011,demersjarzynski2015,debievreetal2016}. Before deriving expressions for $g_1$ and $g_2$ in the high-frequency driving regime, it is worth examining the central role that a separation of timescales plays in our analysis.

We have assumed, after \eqref{eq:smallT}, that $\Delta t$ is much smaller than the timescale $\tau_E(\omega,E_0)$ over which the energy of the system changes significantly.  This condition ensures that the energy increments $\delta E_i$ have approximately identical microcanonical statistics.  We have also assumed that the interval $\Delta t$ contains many subintervals of duration $\delta t$, and that $\delta t >\tau_C(E_0)$, guaranteeing approximate statistical independence among the increments $\delta E_i$.  Thus our analysis involves the hierarchy of timescales:
\begin{equation}
\label{timescales}
T \ll \tau_C(E_0) \ll \Delta t \ll \tau_E(\omega,E_0).
\end{equation}
Since $\tau_E\to\infty$ as $\omega\to\infty$, this hierarchy can be satisfied for any particular energy shell $E_0$ by setting $\omega$ sufficiently large. We conclude that Eq.~\eqref{fp} is valid over an interval of the energy axis whose extent is determined by, and increases with, the value of $\omega$.

The above arguments suggest that the energy diffusion description is valid on a coarse-grained timescale of order $\Delta t$. On shorter timescales, computing the fine details of the system's energy evolution requires the full Hamiltonian equations of motion \eqref{eom}. These details vary greatly from system to system. However, as we will see, the characteristics of the energy diffusion process ultimately depend only on a few key details of these system-specific dynamics, as captured in the coefficients $g_1$ and $g_2$.

\subsection{Drift and diffusion coefficients}
\label{driftdiffusion}

Under \eqref{fp} an initial distribution $\eta (E,0) = \delta (E-E_0)$ evolves after a time $\Delta t\ll\tau_E$ to a distribution $\eta (E,\Delta t)$ with mean and variance \cite{gardiner1985}:
\begin{eqnarray}
\label{eq:mean}
\mathrm{Mean} (E) &=& E_0 + g_1(E_0,\omega) \Delta t \\
\label{eq:var}
\mathrm{Var} (E) &=& g_2(E_0,\omega) \Delta t .
\end{eqnarray}
We can thus determine $g_2$ by calculating $\mathrm{Var} (E)$, the energy spread acquired by an ensemble of trajectories with initial energy $E_0$, evolved for a time $\Delta t$ under the driven Hamiltonian. We perform this calculation in Section \ref{g2calc} of the Appendix, obtaining, in the limit of large $\omega$,
\begin{eqnarray}
\label{g2resub}
g_2(E,\omega) &=& \frac{1}{2} S(\omega;E) \ge 0 \\
\label{eq:spectrumresub}
S(\omega;E) &=& \int_{-\infty}^{\infty} dt \, e^{-i \omega t} C(t;E) \, ,
\end{eqnarray}
where
\begin{equation}
\label{eq:correlationresub}
C(t;E) \equiv \langle \dot{V}(z_0^0) \dot{V}(z_t^0) \rangle - \langle \dot{V}(z_0^0) \rangle \langle \dot{V}(z_t^0) \rangle
\end{equation}
is the microcanonical autocorrelation function of the observable $\dot V(z)$.  Specifically, the averages denoted by $\langle\cdot\rangle$ are computed by sampling initial conditions $z_0^0$ from a microcanonical ensemble at energy $E$, then evolving for time $t$ under $H_0(z)$. By the Wiener-Khinchin theorem \cite{kuboetal2012}, the Fourier transform of $C(t;E)$ is the power spectrum of $\dot{V}(z_t^0)$ at energy $E$, denoted by $S(\omega;E)$. Note that \eqref{g2resub} gives $g_2$ entirely in terms of properties of the \textit{undriven} system, as $C(t;E)$ and thus $S(\omega;E)$ are defined in terms of the undriven trajectories $z_t^0$.

In solving for $g_2$ we approximated driven trajectories $z_t$ by their undriven counterparts $z_t^0$.  As a result, we expect that \eqref{g2resub} contains correction terms that become negligible in the high-frequency limit $\omega\rightarrow\infty$.

In Section \ref{fdproof} of the Appendix we use Liouville's theorem, which expresses the incompressibility of phase space volume under Hamiltonian dynamics \cite{dorfman1999}, to obtain the following expression for the drift coefficient $g_1$ in terms of $g_2(E,\omega)$ and the density of states $\Sigma(E)$ \eqref{eq:dos}:
\begin{equation}
\label{fdresub}
g_1(E,\omega) = \frac{1}{2 \Sigma} \frac{\partial}{\partial E} \Big( g_2 \Sigma \Big).
\end{equation}
This result is a fluctuation-dissipation relation, similar to others previously established for various driven Hamiltonian systems \cite{brownetal-prl1987,brownetal1987,wilkinson1990,jarzynski1992,jarzynski1993,cohen2000,buninetal2011}. 

Using \eqref{g2resub} and \eqref{fdresub}, the Fokker-Planck equation \eqref{fp} takes the compact form
\begin{equation}
\label{eq:compact}
\frac{\partial \eta}{\partial t} = \frac{1}{4} \frac{\partial}{\partial E} \left[ S \Sigma \frac{\partial}{\partial E} \left(\frac{\eta}{\Sigma} \right) \right] .
\end{equation}
Eq.~\eqref{eq:compact} is our main result. It describes the stochastic evolution of the system's energy, under rapid driving, in terms of quantities $S(\omega;E)$ and $\Sigma(E)$ that characterize the undriven system.

As discussed earlier, we expect \eqref{eq:compact} to be valid over a region of the energy axis whose extent depends on $\omega$.  In the next section we assume $\omega$ is sufficiently large that \eqref{eq:compact} is valid over the entire energy axis~\footnote{If $H_0(z)$ has a finite range, then $\omega$ can be chosen so that \eqref{fp} is valid over all allowable energies. If the range of $H_0(z)$ is unbounded then the extent of validity of \eqref{fp} can be made arbitrarily large, though not necessarily infinite, by appropriate choice of $\omega$.}.

\section{Energy absorption and prethermalization}
\label{absorptionprethermal}

\subsection{Energy absorption}
\label{absorption}

We now consider energy absorption, focusing on many-body systems. Under what conditions does the system absorb energy from the rapid drive? Multiplying the Fokker-Planck equation \eqref{fp} by $E$ and  integrating over energy, we obtain
\begin{equation}
\label{avrate}
\frac{d \langle E \rangle}{dt} = \langle g_1 (E,\omega) \rangle,
\end{equation}
where $\langle f \rangle \equiv \int dE \, \eta f$ for any $f(E)$. 
Defining a microcanonical temperature $T_\mu(E)$ via
\begin{equation}
\frac{1}{T_\mu} = \frac{\partial s}{\partial E} \, ,
\end{equation}
where $s(E) \equiv k_B \log \Sigma (E)$ is the microcanonical entropy and $k_B$ is Boltzmann's constant, \eqref{fdresub} becomes:
\begin{equation}
\label{fd3}
g_1(E,\omega) = \frac{1}{2 k_B T_{\mu}} \left[ g_2(E,\omega) + k_B T_\mu \frac{\partial g_2 (E,\omega)}{\partial E} \right].
\end{equation}
The expression in square brackets is an expansion of $g_2 (E + k_B T_\mu, \omega)$ for small $k_B T_\mu$, truncated after first order. For a  system with $N$ degrees of freedom, the difference between $E$ and $E + k_B T_\mu$ corresponds to an energy change of $k_B T_\mu/N$ per degree of freedom. When $N\gg 1$ this change is negligible and \eqref{g2resub}, \eqref{avrate}, \eqref{fd3} give
\begin{equation}
\label{avincrease}
\frac{d \langle E \rangle}{dt} = \Bigl\langle \frac{S(\omega;E)}{4k_BT_\mu}\Bigr\rangle.
\end{equation}

For a many-body system with an unbounded phase space, such as a gas or liquid, the density of states $\Sigma(E)$ increases with energy, hence $T_\mu(E)>0$ and \eqref{g2resub}, \eqref{avincrease} imply that the average energy of the system continually increases with time, as expected intuitively.

If the phase space of the system is bounded, then we expect $T_\mu(E)<0$ at some energies.  For example, for $N$ classical spins described by $H_0 = {\bf B}\cdot \sum_n {\bf S}_n$, $T_\mu(E) < 0$ when $E > 0$.  Thus $d\langle E\rangle/dt$ can be negative.  In this situation we can view the normalized density of states, $\overline{\Sigma} (E) \equiv \Sigma(E)/\int dE' \, \Sigma (E')$, as the ``infinite temperature'' energy distribution, obtained by considering the canonical energy distribution $\eta_{T_c}(E) \propto \Sigma(E) e^{-E/k_B T_c}$ in the limit $T_c\rightarrow\infty$.
If $g_2(E)$ is strictly positive for all $E$, ensuring that there are no insurmountable barriers along the energy axis, then Eq.~\eqref{eq:compact} describes an \textit{ergodic} Markov process, and $\overline{\Sigma} (E)$ is the unique stationary distribution to which any initial distribution evolves as $t\rightarrow\infty$ \cite{gilks1995,gallager2013}.

We thus identify two possible energetic fates of a many-body system in the rapid driving regime. If the phase space is unbounded, then the average energy of the system increases indefinitely, whereas if the system admits a normalized stationary distribution $\overline{\Sigma} (E)$ then the system evolves to this infinite temperature distribution.

\subsection{Prethermalization}
\label{prethermal}

In either case, the energy dynamics predicted by the energy diffusion description relate to the phenomenon of {\it prethermalization}. A driven system is said to prethermalize if it reaches thermal equilibrium with respect to an effective Hamiltonian on short to intermediate timescales, before ultimately gaining energy at far longer times \cite{elseetal2017,abaninetal2017,herrmannetal2017,mori2018,morietal2018,mallayyaetal2019,howelletal2019,machadoetal2019,rajaketal2019,machadoetal2020,rubio-abadal2020,pengetal2021}. In our case, if the system is prepared in a non-microcanonical (i.e.\ non-equilibrium) distribution on a particular energy shell $E_0$, then after a characteristic mixing time the distribution on this energy shell becomes effectively microcanonical, i.e.\ prethermalization occurs with respect to $H_0$, at nearly constant energy.
On longer timescales, the energy dynamics are governed by the Fokker-Planck equation \eqref{eq:compact}, and the system absorbs energy from the drive $V(t)$ \eqref{avincrease}.
For large $\omega$ this absorption can be exceedingly slow, as the power spectrum $S(\omega;E)$ decays faster than any power of $\omega^{-1}$ for any smooth $H_0$ \cite{bracewell1978}.
This is consistent with observed exponential-in-frequency suppression of energy absorption in a range of classical and quantum model systems \cite{abaninetal2015,morietal2016,kuwaharaetal2016,elseetal2017,abaninetal2017,mori2018,howelletal2019,machadoetal2019,tranetal2019,rubio-abadal2020,pengetal2021}. Prethermalization thus occurs when $\omega$ lies deep within the tail of the power spectrum.


Following the above-mentioned initial relaxation, energy absorption is slow but does not vanish. As the system energy $E$ gradually grows, the intrinsic correlation time $\tau_C(E)$ generically decreases with increasing particle velocities, hence the power spectrum $S(\omega;E)$ broadens. Eventually, at sufficiently large $E$, the drive frequency might no longer be located in the far tail of the power spectrum: This marks the onset of unsuppressed energy absorption toward the infinite-temperature state. If the phase space is bounded, then $\omega$ can be chosen so that energy absorption is suppressed on all energy shells; in this case energy absorption remains very slow throughout the system's evolution towards the infinite-temperature energy distribution $\overline{\Sigma} (E)$.

Energy absorption from periodic driving has also been studied using the Floquet-Magnus (FM) expansion, which, for time-periodic $H(z,t)$, expresses the associated ``Floquet'' Hamiltonian $H_F(z)$ as a perturbative expansion in powers of $\omega^{-1}$.  $H_F$ is a time-\textit{independent} Hamiltonian whose dynamics coincide with those of $H(t)$ at stroboscopic times $t=0,T,2T...$.  At high frequencies and short timescales, the evolution obtained by truncating the FM expansion at some order is expected to be a good approximation of the exact dynamics \cite{rahavetal2003prl,rahavetal2003,bukovetal2015,abaninetal2017,higashikawaetal2018}.  See e.g.\ the fourth-order (in $\omega^{-1}$) expression for $H_F(z)$ derived in \cite{rahavetal2003prl,rahavetal2003} for a system in one degree of freedom.  By contrast it appears that our results are not obtainable via the FM expansion. For smooth $H_0$, the coefficient $g_2(E,\omega)$ decays faster than any power of $\omega^{-1}$ at large $\omega$ (as mentioned earlier), and thus cannot be described accurately by an FM-like expansion in powers of $\omega^{-1}$. Indeed, this might have been anticipated, as high frequency driving cannot induce unbounded energy absorption unless the FM expansion diverges \cite{bukovetal2015,ponteetal2015a,abaninetal2017,morietal2018}.

\section{Quantum-classical correspondence}
\label{quantum}

Energy absorption, prethermalization, and relaxation to the infinite temperature state have been documented for a variety of periodically driven quantum systems \cite{buninetal2011,lazaridesetal2014,dalessiorigol2014,ponteetal2015a,rehnetal2016,abaninetal2017}.
It is instructive to ask how the classical energy diffusion described by \eqref{eq:compact} might emerge, in agreement with the correspondence principle, as the semiclassical limit of quantum dynamics. We now briefly describe a model that illustrates this correspondence; similar analyses may be found elsewhere in the literature on energy diffusion \cite{cohen2000,elyutin2006}.

Consider a quantum system governed by a Hamiltonian $\hat H_0 + \hat V \cos(\omega t)$, the counterpart of \eqref{hamiltonian}.
Let us model the system's evolution as a random walk in the spectrum of $\hat H_0$, with stochastic quantum ``jumps'' from one energy level to another.
By Fermi's golden rule, the transition rate from energy $E$ to $E \pm \hbar \omega$ is given by
\begin{equation}
\Gamma_{\pm}=\frac{\pi}{2 \hbar} \overline{|V_{m n}|^2}\rho(E_n) \, ,
\end{equation}
where $V_{m n} = \langle m\vert\hat V\vert  n\rangle$ is the matrix element of $\hat{V}$ associated with the energy levels $E_m$ and $E_n$ of $\hat{H}_0$; the overbar denotes an average over a narrow range of matrix elements with $E_m \approx E$ and $E_n \approx E \pm \hbar \omega$; and $\rho(E) = \Sigma(E)/h^N$ is the semiclassical density of states. As $\hbar \to 0$, the spectrum of $\hat H_0$ becomes dense and our random walk model leads naturally to a description in terms of energy diffusion, with drift and diffusion coefficients
\begin{equation}
\label{eq:qcoeffs}
g_1 = (\Gamma_+ - \Gamma_-)(\hbar \omega) \quad, \quad g_2 = (\Gamma_+ + \Gamma_-)(\hbar \omega)^2.
\end{equation}

A semiclassical estimate for matrix elements of quantized chaotic systems \cite{feingoldperes1986,wilkinson1987} gives
\begin{equation}
\label{semiclassical}
\overline{|V_{m n}|^2} \approx h^{N-1} \frac{S_V(\omega;\overline{E})}{\Sigma (\overline{E})} \, ,
\end{equation}
where $\overline{E} \equiv (E_m+E_n)/2$. Here, $S_V(\omega;E)$ is the power spectrum for the classical observable $V$, and is related to $S(\omega;E)$ (the power spectrum for $\dot V$) via $S = \omega^2 S_V$. Combining results, we find that \eqref{eq:qcoeffs} converges to the classical results \eqref{fdresub} and \eqref{g2resub} as $\hbar\rightarrow 0$. While this analysis is based on a heuristic model that ignores quantum coherences, it suggests that our classical energy diffusion picture is relevant for understanding periodically driven quantum systems; in particular it provides a semiclassical explanation for the observed exponential-in-frequency suppression of energy absorption \cite{abaninetal2015,morietal2016,kuwaharaetal2016,elseetal2017,abaninetal2017,machadoetal2019,tranetal2019,rubio-abadal2020,pengetal2021}.

\section{Conclusion}
\label{conclusion}

We have analyzed the diffusive energy dynamics of chaotic, ergodic Hamiltonian systems under rapid periodic driving. Observing that the system's dynamics are only weakly affected by very rapid driving, we have established a Fokker-Planck equation governing the evolution of the system's energy probability distribution. Our analysis predicts a generic, long-lived prethermal state, and for many-body systems our results point to two possible energetic fates: indefinite energy growth, or relaxation to the infinite-temperature equilibrium state. In the semiclassical limit, a model of energy absorption for periodically driven, quantized chaotic systems coincides with our purely classical energy diffusion description.

A central feature of our Fokker-Planck equation is that the drift and diffusion coefficients $g_1$ and $g_2$ are determined by the undriven dynamics.  A similar situation arises in linear response theory (LRT), where transport coefficients, such as electrical conductivities, in a system subject to {\it weak} time-periodic driving, are expressed in terms of correlation functions computed in the absence of driving~\cite{dorfman1999}.  In LRT these results are obtained perturbatively, through a formal expansion in powers of the driving strength.  It is unclear whether our results can similarly be obtained through a perturbative  expansion.  A natural candidate for a small parameter in our case is the inverse frequency $\omega^{-1}$, but this seems to lead to the Floquet-Magnus expansion, which as already noted at the end of Sec.~\ref{prethermal} is somewhat at odds with our analysis.  Both this discrepancy, and the question of whether our results can be obtained through a formal perturbative expansion, bear further investigation.

Low-dimensional billiard systems -- in which a particle in a cavity alternates between straight-line motion and specular reflection off the cavity walls -- offer an ideal testing ground for the theory presented in this paper, as certain billiard shapes are rigorously proven \cite{sinai1970,bunimovich1979,wojtkowski1986} to generate chaotic, ergodic motion.
Energy absorption in driven billiard systems, sometimes known as \textit{Fermi acceleration}, is a well-studied phenomenon \cite{fermi1949,ulam1961,jarzynski1993,barnettetal2001,batistic2014,demersjarzynski2015}, although much of the existing literature focuses on the case of slow driving.
In a forthcoming work, we will present numerical evidence for the validity of the Fokker-Planck equation \eqref{eq:compact} for a particle in a chaotic billiard subject to a spatially uniform, rapidly time-periodic force.  For this system, the driven and undriven trajectories of the particle can be computed to machine precision and \eqref{eq:compact} can be solved analytically, allowing for an especially precise test of the energy diffusion description.

 Our results may  also be tested for  previously studied many-body classical systems, such as a many-body generalization of the kicked rotor model \cite{chirikov1979,ott2002} that exhibits unbounded energy absorption in a range of parameter regimes \cite{rajaketal2018,notarnicolaetal2018,rajaketal2019}.  Energy absorption has also been studied in the classical driven Heisenberg spin chain \cite{mori2018,khassehetal2019,howelletal2019}. For these models and others, we expect our analysis to apply only if the time-averaged Hamiltonian $H_0$ generates chaotic and ergodic dynamics.

\section*{Acknowledgements}

We gratefully acknowledge stimulating discussions with Ed Ott, David Levermore, and Saar Rahav, and financial support from the DARPA DRINQS program (D18AC00033).

\renewcommand{\theequation}{A\arabic{equation}}
\setcounter{equation}{0}
\setcounter{subsection}{0}

\appendix
\section*{Appendix}

Here, we obtain an expression for the energy diffusion coefficient $g_2$, given by \eqref{g2resub}. We then derive the fluctuation-dissipation relation \eqref{fdresub}.

\subsection{Calculation of $g_2$}
\label{g2calc}

We begin with relation \eqref{eq:var}. According to this equation, calculating $g_2$ amounts to computing $\mathrm{Var} (E)$, the variance in energy acquired by an ensemble of trajectories with initial energy $E_0$, evolved for a time $\Delta t$ under the driven Hamiltonian. Specifically, we consider an ensemble of driven trajectories evolving from microcanonically sampled initial conditions at $t=0$. Upon integrating \eqref{power} along these trajectories, we obtain (with no approximations so far)
\begin{equation}
\label{var2}
\mathrm{Var}(E) =  \int_0^{\Delta t} \int_0^{\Delta t} dt \, dt' \cos (\omega t) \cos (\omega t') C_{neq}(t,t';E_0) ,
\end{equation}
where
$C_{neq}(t,t';E_0) \equiv \langle \dot{V}(z_t) \dot{V}(z_{t'}) \rangle - \langle \dot{V}(z_t) \rangle \langle \dot{V}(z_{t'}) \rangle$
is a nonequilibrium correlation function and angular brackets $\langle\cdot\rangle$ denote an ensemble average.
In the high-frequency limit $\omega\to\infty$, as driven trajectories $z_t$ approach their undriven counterparts $z_t^0$, $C_{neq}(t,t';E_0)$ can be replaced by the equilibrium correlation function

\begin{equation}
C(t'-t;E_0) \equiv \langle \dot{V}(z_t^0) \dot{V}(z_{t'}^0) \rangle - \langle \dot{V}(z_t^0) \rangle \langle \dot{V}(z_{t'}^0) \rangle ,
\end{equation}

\noindent which depends only on the difference $t'-t$, due to the time-translation symmetry of the microcanonical distribution under the undriven dynamics.

Replacing $C_{neq}(t,t';E_0)$ by $C(t^\prime-t;E_0)$ in \eqref{var2}, and using standard manipulations to evaluate the double integral (see, e.g. \cite{reif}), we arrive at
\begin{equation}
\label{var3}
\mathrm{Var}(E) \approx \frac{1}{2} S(\omega;E_0) \Delta t ,
\end{equation}
where
\begin{equation}
S(\omega;E_0) = \int_{-\infty}^{\infty} dt \, e^{-i \omega t} C(t;E_0)
\end{equation}
is the power spectrum of $\dot V(z_t^0)$, which is equal to the Fourier transform of $C(t;E_0)$ by the Wiener-Khinchin theorem \cite{kuboetal2012}.  The approximation in \eqref{var3} contains correction terms that are sublinear in $\Delta t$. Comparing \eqref{var3} with \eqref{eq:var} and relabeling $E_0$ as $E$, we obtain \eqref{g2resub}, our final expression for $g_2$.

\subsection{Calculation of $g_1$}
\label{fdproof}

We now derive \eqref{fdresub}, which expresses a fluctuation-dissipation relation between the drift and diffusion coefficients $g_1$ and $g_2$. To do so, we first note that the constant function $\rho(z) = 1$ is a stationary solution to the Liouville equation \eqref{liouville}. This reflects the incompressibility of phase space volume under Hamiltonian dynamics (Liouville's theorem) \cite{dorfman1999}. Since $\rho = 1$ is stationary under the dynamics in \textit{phase space}, the corresponding (unnormalized) distribution in \textit{energy space} should be stationary under the Fokker-Planck equation. This energy distribution, obtained by marginalizing over the constant solution $\rho = 1$, is the density of states $\Sigma(E)$ -- see \eqref{eq:dos}.
Setting $\eta(E,t) = \Sigma (E)$ as a stationary solution of the Fokker-Planck equation \eqref{fp}, we have
\begin{equation}
\label{stationary}
0 = 
 - \frac{\partial}{\partial E} \left[ g_1 \Sigma - \frac{1}{2} \frac{\partial}{\partial E} \left( g_2 \Sigma \right) \right] .
\end{equation}
Thus the quantity in square brackets is constant as a function of $E$. We label this constant by $\alpha$:

\begin{equation}
\label{eq:JSigma}
\alpha \equiv g_1 \Sigma - \frac{1}{2} \frac{\partial}{\partial E} \left( g_2 \Sigma \right).
\end{equation}

\noindent We now aim to show that $\alpha = 0$, which then immediately implies the fluctuation-dissipation relation \eqref{fdresub}.

To proceed, we first use \eqref{eq:JSigma} to eliminate $g_1$ from the Fokker-Planck equation \eqref{fp},
obtaining
\begin{equation}
\label{eq:fp-new}
\frac{\partial\eta}{\partial t} = -\alpha \frac{\partial}{\partial E}\left(\frac{\eta}{\Sigma}\right)
+ \frac{1}{2} \frac{\partial}{\partial E}\left[ g_2\Sigma \frac{\partial}{\partial E}\left(\frac{\eta}{\Sigma}\right) \right] .
\end{equation}

In the main text, in arguing that the system energy evolves diffusively, we considered trajectories with a common initial energy $E_0$, and we arrived at the hierarchy of timescales \eqref{timescales} required for the validity of the energy diffusion picture: $T \ll \tau_C(E_0) \ll \Delta t \ll \tau_E(\omega,E_0)$.
Since $\tau_E\rightarrow\infty$ as $\omega\rightarrow\infty$, this  hierarchy suggests that for a given, sufficiently large value of $\omega$, there is a range of energies over which \eqref{eq:fp-new} is valid.
This range can be enlarged by increasing the value of $\omega$, but there might exist no value $\omega^*$ such that \eqref{eq:fp-new} is valid over the entire energy axis for all $\omega>\omega^*$.
Thus let us fix the value of $\omega$ and let $[a,b]$ denote a finite interval of the energy axis, such that \eqref{eq:fp-new} is valid for energies $a\le E\le b$.
The existence of such an interval is sufficient to establish that $\alpha=0$, as we now show.

\begin{figure}[tbp]
\includegraphics[trim = .5in 0in 0in 0in , scale=0.4,angle=0]{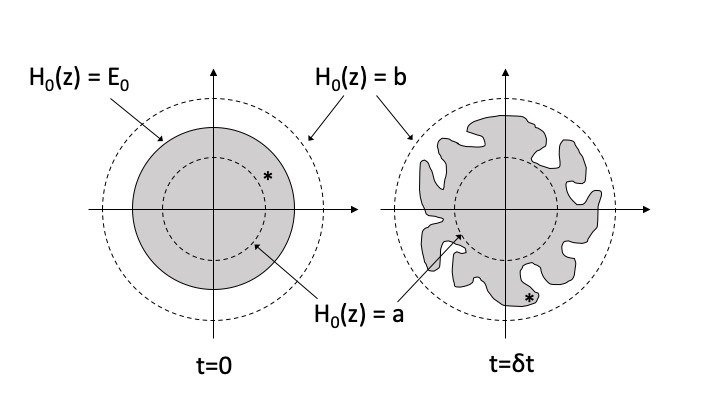}
\caption{
Schematic depiction of phase space.
On the left, the initial distribution is uniform, $\rho(z,0)=c$ (shaded), up to a cutoff energy $E_0$.
The figure on the right shows the distribution a short time later, after evolution under $H(z,t)$;
$\rho(z,\delta t)=c$ in the shaded region.
Only in the annular region between the two energy shells $a$ and $b$ (dashed circles) does $\rho(z,t)$ vary with time for $t \in [0,\delta t]$.
Asterisks depict initial and final conditions for a representative trajectory.
}
\label{fig}
\end{figure}

Consider an ensemble of trajectories evolving under $H(z,t)$, from an initial phase space distribution that is uniform up to a cutoff $E_0 \in (a,b)$:
\begin{equation}
\label{eq:init}
\rho(z,0) = c\,\theta(E_0-H_0(z)),
\end{equation}
where $\theta(\cdot)$ is the unit step function and $c^{-1} = \Omega(E_0)$ is the volume of phase space enclosed by the energy shell $E_0$ (which was assumed finite in Section \ref{diffusion}).
The corresponding energy distribution is
\begin{eqnarray}
\eta(E,0) &=& \int dz \, \delta(E-H_0) \rho(z,0) \nonumber \\
&=&c \, \Sigma(E) \theta(E_0-E) .
\end{eqnarray}

As this ensemble of trajectories evolves in time, the value of the density $\rho$ at any $(z,t)$ is either $c$ or 0, by Liouville's theorem.
For a sufficiently short but finite interval $0 \le t \le \delta t$, $\rho(z,t)$ remains constant outside the region of phase space between the two energy shells $H_0=a$ and $H_0=b$ (see Fig.~\ref{fig}), hence
\begin{equation}
\label{eq:Liouville}
\eta(E,t) = 
\begin{cases}
c\,\Sigma(E) & \text{if $E \le a$} \\
0 & \text{if $E \ge b$}
\end{cases}
\end{equation}
We emphasize that \eqref{eq:Liouville} is exact, and a direct consequence of Liouville's theorem.

\eqref{eq:Liouville} implies that for $t\in[0,\delta t]$ there is no net flow of probability into or out of the energy interval $[a,b]$:
\begin{equation}
\label{eq:zero}
0 = \frac{d}{dt}
\int_a^b \eta(E,t) \, dE .
\end{equation}
We can use \eqref{eq:fp-new}, which is valid in $[a,b]$, along with \eqref{eq:Liouville} to evaluate the right side of \eqref{eq:zero}, obtaining
\begin{equation}
\label{alphazero}
0 =
\left[ - \alpha \left(\frac{\eta}{\Sigma}\right) 
+ \frac{1}{2} g_2\Sigma \frac{\partial}{\partial E}\left(\frac{\eta}{\Sigma}\right) \right]_a^b 
= \alpha c ,
\end{equation}
hence $\alpha=0$. By \eqref{eq:JSigma}, this establishes the fluctuation-dissipation relation \eqref{fdresub}.

\bibliography{refs}{}

\end{document}